\documentclass{emulateapj}
\usepackage{apjfonts}
\usepackage{lscape}

\shorttitle{ULP Cepheids}
\shortauthors{Bird, Stanek, \& Prieto}

\newcommand{\eg}{{\rm e.g.}}

\newcommand{\osmc}{OGLE Cepheid}
\newcommand{\pls}{PL$_{SMC}$}
\newcommand{\plu}{PL$_{ULP}$}

\begin{document}

\title{Using Ultra Long Period Cepheids to Extend the Cosmic Distance
Ladder to 100 Mpc and Beyond}

\author{Jonathan C. Bird, K. Z. Stanek, Jos\'{e} L. Prieto}
\affil{Department of Astronomy, The Ohio State University, 140 West
18th Avenue, Columbus, OH 43210, bird@astronomy.ohio-state.edu,
kstanek@astronomy.ohio-state.edu, prieto@astronomy.ohio-state.edu}

\slugcomment{Accepted for publication in ApJ.}
\begin{abstract}

We examine the properties of 18 long period ($80-210$ days) and very
luminous (median absolute magnitude of M$_I=-7.86$ and M$_V=-6.97$)
Cepheids to see if they can serve as an useful distance indicator. We
find that these Ultra Long Period (ULP) Cepheids have a relatively
shallow Period-Luminosity (PL) relation, so in fact they are more
``standard candle''-like than classical Cepheids. In the
reddening-free Wesenheit index, the slope of the ULP PL relation is
consistent with zero. The scatter of our sample about the W$_I$ PL
relation is $0.23$ mag, approaching that of classical Cepheids and
Type Ia Supernovae. We expect this scatter to decrease as bigger and
more uniform samples of ULP Cepheids are obtained. We also measure a
non-zero period derivative for one ULP Cepheid (SMC HV829) and use the
result to probe evolutionary models and mass loss of massive
stars. ULP Cepheids main advantage over classical Cepheids is that
they are more luminous, and as such show great potential as stellar
distance indicators to galaxies up to $100$ Mpc and beyond.
\end{abstract}

\keywords{Cepheids --- stars: distances --- stars: mass loss --- distance scale}

\section{Introduction}\label{sec:intro}

A reliable method of measuring the physical distance to astrophysical
objects has always been sought after in observational astronomy
\citep[\eg,][]{Bessel1839}. In the era of ``precision cosmology'', the
need for accurate physical distance measurements has been amplified
\citep[\eg,][]{Spergel03, Riess04, Tegmark04}. Accurate and precise
distance indicators hold the key to pinning down the value of the
Hubble constant $(H_0)$ and many other cosmological parameters
\citep[see discussion in, \eg,][]{Macri06}. A number of methods have
been employed to determine extragalactic distances, with varying
degree of success \citep[\eg,][]{Freedman01}. The construction and
reliability of the ``cosmological distance ladder'' depends crucially
on Cepheid variables being able to provide precise and accurate
distances to nearby ($d\lesssim20\;$Mpc) galaxies.  The quest for such
distances has been an arduous journey for almost a hundred years, with
many dramatic twists and turns \citep[for a review of early years,
see][for a recent review see, \eg, \citealt{Macri05}]{Baade1956}.
 
Cepheids offer several advantages as distance indicators. Massive
stars $(\ge5 M_{\odot})$ make an excursion through the instability
strip and most, if not all, of them become Cepheid variables. These
variable stars are relatively bright ($M_V\sim -4$ for a $P\sim
10\;$day Cepheid) and often have large brightness variations
(amplitude $\sim1$ mag) with a characteristic ``saw-tooth'' shaped
light curve. Their intrinsic brightness, combined with their light
curve shape and colors, make them easy to distinguish from other
classes of variable stars. As a result, Cepheids have been detected
and studied in a significant number of star-forming galaxies. The
physical mechanisms underlying Cepheid pulsation are well understood,
including the observed tight period-luminosity (PL) relationship
\citep[\eg,][]{Chiosi93}. The small scatter in the PL relation allows
distance measurements precise to $\sim 5\%$ \citep[\eg,][]{Macri06}.
For these reasons, Cepheids are commonly used to calibrate other
distance indicators, forming the base of the cosmological distance
ladder.

Despite their many advantages as a distance indicator, Cepheid
distances also have some shortcomings. Most Cepheids have an intrinsic
brightness of $M_V \ge -5$, so with the current instrumentation they
can be only used to measure distances to $\lesssim30\;$Mpc (the
largest Cepheid distance in \citealt{Freedman01} is $\sim
22\;$Mpc). Observations of Cepheids in distant galaxies are also
hindered by blending \citep{Mochejska00}--- as young stars, Cepheids
live in close proximity to the crowded star-forming regions of their
host galaxies, and are thus likely to have another star of similar
brightness on the scale of a typical instrumental
point-spread-function (PSF). The effect of blending becomes worse as
the square of the distance to the host galaxy \citep{Stanek99}, again
limiting the usefulness of Cepheids to measuring distances
$\lesssim30\;$Mpc even with high resolution instruments such as the
Hubble Space Telescopes ({\em HST}). Ideally, we would like to find a
distance indicator that shares the good properties of classical
Cepheids, but is even more luminous, allowing us to observe it further
away and be less susceptible to blending. In this paper we discuss
such a possible distance indicator, namely Ultra Long Period (ULP)
Cepheids.

We define ULP Cepheids as fundamental mode Cepheids with pulsation
periods greater than 80 days. Several such Cepheids have been already
discovered in the pioneering study of \citet{Leavitt1908}. However,
ULP Cepheids have traditionally been ignored for distance measurements
as they are PL outliers.  Indeed, the observed PL relation flattens
for Cepheids with periods greater than 100 days
\citep[\eg,][]{Grieve85,Freedman92}. \citet{Grieve85} suggests that
long period Leavitt Variables could be used for distance measures---
unfortunately that idea has not permeated through the community.  We
argue that the flattening of the PL at long periods actually improves
the usefulness of ULP Cepheids as distance indicators because it makes
them a good standard candle in the traditional sense. We note several
additional advantages of ULP Cepheids over lower period Cepheids due
to their increased luminosity. ULP Cepheids could be used as a stellar
distance measure to the Hubble Flow (up to $\sim150$ Mpc)--- several
times the current observational limit. In Section~\ref{sec:sample} we
describe our sample compiled from the literature. The ULP Cepheid PL
relation is discussed in
Section~\ref{sec:DM}. Section~\ref{sec:massloss} demonstrates how ULP
Cepheids may provide the additional benefit of testing massive stellar
evolutionary models. We summarize our results in
Section~\ref{sec:conclusion}.

\section{Sample}\label{sec:sample}

\begin{figure*}
\figurenum{1}
\plotone{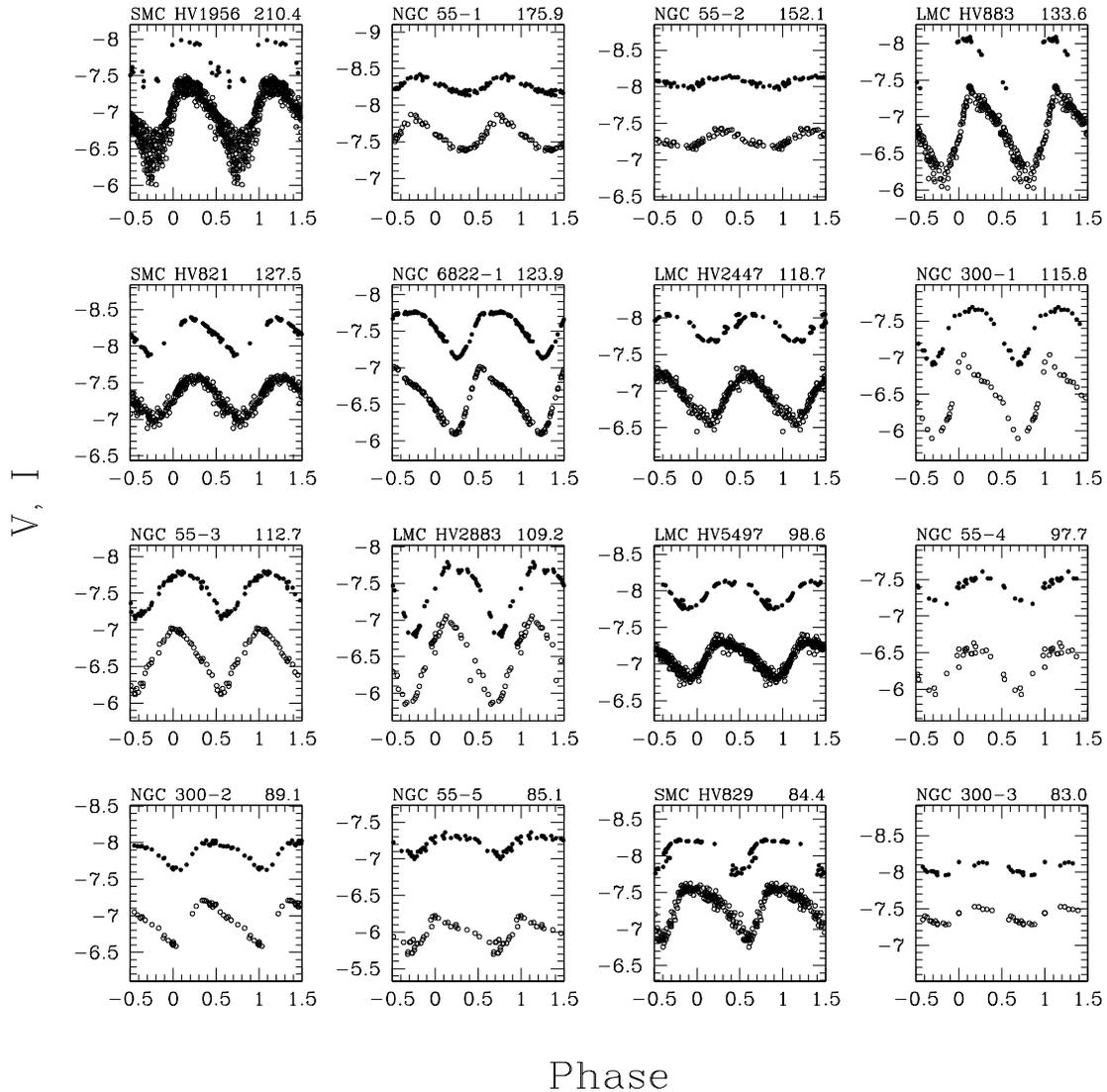}
\caption{\label{fig:lcs}The $V$ (open circles) and $I$ (filled
circles, where available) band light curves of the ULP Cepheids. Each
panel spans 2.4 magnitudes along the y-axis. The phase is given along
the x-axis. The Cepheid identification (see column 1,
Table~\ref{tab:ceps}) is listed in the upper left of each plot while
the period (in days) is in the upper right.}
\end{figure*}

\tabletypesize{\scriptsize}
\begin{deluxetable*}{llccccccccccccc}
\tablewidth{0pt}
\tablecolumns{15}
\setlength{\tabcolsep}{0.005in} 
\tablecaption{Ultra Long Period Cepheids}
\tablehead{
\colhead{ID} &
\colhead{Host Galaxy} &
\colhead{RA} &
\colhead{DEC} &
\colhead{P} &
\colhead{$<V>$} &
\colhead{$(V-I)$} &
\colhead{W$_I$} &
\colhead{$E(B-V)$} &
\colhead{(m-M)$_0$} &
\colhead{$V_0$} &
\colhead{$(V-I)_0$} &
\colhead{$W_{I0}$} &
\colhead{$12 + \log($O/H$)$} &
\colhead{References} \\
\colhead{} &
\colhead{} &
\colhead{(J2000.0)} &
\colhead{(J2000.0)} &
\colhead{(days)} &
\colhead{(mag)} &
\colhead{(mag)} &
\colhead{(mag)} &
\colhead{(mag)} &
\colhead{(mag)} &
\colhead{(mag)} &
\colhead{(mag)} &
\colhead{(mag)} &
\colhead{dex} &
\colhead{}  \\
\colhead{(1)} &
\colhead{(2)} &
\colhead{(3)} &
\colhead{(4)} &
\colhead{(5)} &
\colhead{(6)} &
\colhead{(7)} &
\colhead{(8)} &
\colhead{(9)} &
\colhead{(10)} &
\colhead{(11)} &
\colhead{(12)} &
\colhead{(13)} &
\colhead{(14)} &
\colhead{(15)} \\
}

\startdata
LMC HV883 &   LMC &       $05:00:08.6$ &	$-68:27:03$ &	$133.6$ &	$12.12$ &	$1.09$ &	$9.34$ &	$0.14$ &	$18.50$ &	$-6.83$ &	$0.91$ &	$-9.16$ &	$8.39\pm0.12$ & 1,2,3  \\
LMC HV2447 &  LMC &       $05:19:31.4$ &	$-68:41:12$ &	$118.7$ &	$11.99$ &	$1.12$ &	$9.13$ &	$0.14$ &	$18.50$ &	$-6.96$ &	$0.94$ &	$-9.37$ &	$8.39\pm0.12$ & 1,2,3  \\
LMC HV2883 &  LMC &       $04:56:27.0$ &	$-64:41:26$ &	$109.2$ &	$12.41$ &	$1.07$ &	$9.68$ &	$0.14$ &	$18.50$ &	$-6.54$ &	$0.89$ &	$-8.82$ &	$8.39\pm0.12$ & 1,2,3  \\
LMC HV5497 &  LMC &       $04:55:40.0$ &	$-66:25:39$ &	$98.6$ &	$11.92$ &	$1.11$ &	$9.09$ &	$0.14$ &	$18.50$ &	$-7.03$ &	$0.93$ &	$-9.41$ &	$8.39\pm0.12$ & 1,2,3  \\
SMC HV1956 &  SMC &	  $01:04:15.9$ &	$-72:45:20$ &	$210.4$ &	$12.28$ &	$0.83$ &	$9.95$ &	$0.09$ &	$18.93$ &	$-6.94$ &	$0.71$ &	$-8.98$ &	$7.98\pm0.10$ & 5,2,3;4  \\
SMC HV821 &   SMC &       $00:41:43.5$ &	$-73:43:24$ &	$127.5$ &	$11.92$ &	$1.03$ &	$9.29$ &	$0.09$ &	$18.93$ &	$-7.30$ &	$0.92$ &	$-9.64$ &	$7.98\pm0.10$ & 1,6,3;4  \\
SMC HV829 &   SMC &       $00:50:28.9$ &	$-72:45:09$ &	$84.4$ &	$11.97$ &	$0.91$ &	$9.65$ &	$0.09$ &	$18.93$ &	$-7.25$ &	$0.80$ &	$-9.28$ &	$7.98\pm0.10$ & 1,6,3;4  \\
NGC 6822-1 &  NGC 6822 &  $19:45:02.0$ &	$-14:47:33$ &	$123.9$ &	$17.86$ &	$1.40$ &	$14.29$ &	$0.36$ &	$23.31$ &	$-6.60$ &	$0.94$ &	$-9.02$ &	$8.11\pm0.10$ & 7,8,9  \\
NGC 55-1 &    NGC 55 &    $00:14:13.0$ &	$-39:08:42$ &	$175.9$ &	$19.25$ &	$0.84$ &	$17.11$ &	$0.13$ &	$26.43$ &	$-7.60$ &	$0.68$ &	$-9.33$ &	$8.05\pm0.10$ & 10,11,12  \\
NGC 55-2 &    NGC 55 &    $00:15:12.0$ &	$-39:12:18$ &	$152.1$ &	$19.56$ &	$0.95$ &	$17.14$ &	$0.13$ &	$26.43$ &	$-7.28$ &	$0.79$ &	$-9.29$ &	$8.05\pm0.10$ & 10,11,12  \\
NGC 55-3 &    NGC 55 &    $00:14:36.6$ &	$-39:11:09$ &	$112.7$ &	$20.18$ &	$1.05$ &	$17.51$ &	$0.13$ &	$26.43$ &	$-6.67$ &	$0.88$ &	$-8.92$ &	$8.05\pm0.10$ & 10,11,12  \\
NGC 55-4 &    NGC 55 &    $00:15:14.3$ &	$-39:13:17$ &	$97.7$ &	$20.54$ &	$1.25$ &	$17.35$ &	$0.13$ &	$26.43$ &	$-6.31$ &	$1.08$ &	$-9.08$ &	$8.05\pm0.10$ & 10,11,12  \\
NGC 55-5 &    NGC 55 &    $00:15:10.1$ &	$-39:12:26$ &	$85.1$ &	$20.84$ &	$1.38$ &	$17.32$ &	$0.13$ &	$26.43$ &	$-6.01$ &	$1.22$ &	$-9.12$ &	$8.05\pm0.10$ & 10,11,12  \\
NGC 300-1 &   NGC 300 &   $00:55:11.6$ &	$-37:33:55$ &	$115.8$ &	$20.13$ &	$0.97$ &	$17.66$ &	$0.10$ &	$26.37$ &	$-6.55$ &	$0.85$ &	$-8.71$ &	$8.25\pm0.22$ & 13,14,15  \\
NGC 300-2 &   NGC 300 &   $00:54:35.0$ &	$-37:35:01$ &	$89.1$ &	$19.71$ &	$1.02$ &	$17.12$ &	$0.10$ &	$26.37$ &	$-6.97$ &	$0.89$ &	$-9.25$ &	$8.25\pm0.22$ & 13,14,15  \\
NGC 300-3 &   NGC 300 &   $00:54:54.3$ &	$-37:37:02$ &	$83.0$ &	$19.26$ &	$0.77$ &	$17.30$ &	$0.10$ &	$26.37$ &	$-7.42$ &	$0.65$ &	$-9.07$ &	$8.25\pm0.22$ & 13,14,15  \\
I Zw 18-1 &    I Zw 18 &  \nodata      &	\nodata &	$129.8$ &	$23.56$ &	$0.74$ &	$21.67$ &	$0.03$ &	$31.30$ &	$-7.84$ &	$0.70$ &	$-9.63$ &	$7.21\pm0.10$ & 16,17\\
I Zw 18-2 &    I Zw 18 &  \nodata      &	\nodata &	$125.0$ &	$23.47$ &	$0.91$ &	$21.15$ &	$0.03$ &	$31.30$ &	$-7.93$ &	$0.87$ &	$-10.15$ &	$7.21\pm0.10$ & 16,17\\
\enddata													        

\tablecomments{\label{tab:ceps}ULP Cepheids in our sample grouped by
host galaxy. (1): Cepheid Identification. (2): Host galaxy
name. (3,4): Right Ascension and Declination in J2000
coordinates. (5): Period in days. (6): Apparent mean $V$ magnitude of
Cepheid. (7): Apparent $(V-I)$ color of Cepheid. (8): Apparent
Wesenheit magnitude (definition in text). (9): Reddening towards host
galaxy. (10): Reddening-free distance modulus. (11): Absolute $V$
magnitude. (12): Absolute $(V-I)$ color. (13): Absolute Wesenheit
magnitude. (14): Metallicity: $12 + \log($O/H$)$. (15): First
reference is for photometry, reddening, and distance modulus (except
where noted in text). Second reference refers to metallicity. If third
reference is present, its reddening and distance modulus measurements
supercedes the first.}

\tablerefs{
References:
1.~\citet{Freedman85}; 
2.~\citet{Pagel78}; 
3.~\citet{Udalski99};
4.~\citet{Hilditch05} and \citet{Keller06};
5.~\citet[][ASAS survey]{Pojmanski97};
6.~\citet{Peimbert76};
7.~\citet{Pietrzynski04}; 
8.~\citet{Peimbert05};
9.~\citet{Gieren06} 
10.~\citet{Pietrzynski06}; 
11.~\citet{Tullmann03}; 
12.~\citet{Gieren08}
13.~\citet{Gieren04}; 
14.~\citet{Urbaneja05};
15.~\citet{Gieren05};
16.~\citet{Fiorentino08};
17.~\citet{Skillman93}.
}
\end{deluxetable*}

We have assembled a sample of ULP Cepheids from the literature and
list their reported positions, periods, and mean $V$ and $I$
magnitudes (see Table~\ref{tab:ceps}). We adopt the periods, reddening
values, and distance moduli found in these sources except in the case
of the Magellanic Clouds (see below). Our primary criteria for
selecting the sample was the existence of $V$ and $I$ data calibrated
on the standard Johnson/Kron-Cousins magnitude system using Landolt
standards (with the possible exception of the Magellanic Clouds; see
below). The ULP distinction is applied to fundamental mode Cepheids
with periods greater 80 days. We combed the recent literature for
reports of such variable stars.

\textbf{Magellanic Clouds} Our sample includes four LMC and three SMC
ULP Cepheids. \citet{Freedman85} calibrated photoelectric observations
of these Cepheids and transformed them to the Johnson/Kron-Cousins
standard system. The mean flux-weighted photometry for the six
Cepheids reported in \citet{Freedman85} agrees with the Landolt
standard star calibrated results of \citet{Moffett98} to within $0.04$
mag, suggesting that the standard photometric system calibration is
robust. Mean flux-weighted photometry for HV1956 is obtained from the
All Sky Automated Survey \citep[ASAS;][]{Pojmanski97} and
\citet{Moffett98}. The $V$ light curves of six of these ULP
Cepheids were obtained from ASAS. HV2883 was not targeted by ASAS,
and its $V$ light curve photometry was obtained from \citet{Madore75},
\citet{vanGenderen83}, and \citet{Moffett98}. \citet{Moffett98}
provide the $I$ light curve data for the entire sample. We applied the
analysis of variance technique \citep{Schwarzenberg89} to the seven
Harvard Variable light curves in Figure~\ref{fig:lcs} to obtain the
periods listed in Table~\ref{tab:ceps}. We adopt total reddening
values of $E(B-V)=0.14$ mag and $E(B-V)=0.09$ mag for the LMC and SMC,
respectively \citep{Udalski99}. We assume a distance modulus (DM) of
$(m-M)_0=18.5$ mag to the LMC for consistency with the sources listed
in Table~\ref{tab:ceps}. The SMC DM used is $(m-M)_0=18.93$ mag
\citep{Hilditch05, Keller06}. The LMC (SMC) hosts ULP Cepheids with
periods of 98.6, 109.2, 118.7, and 133.6 (84.4, 127.5, and 210.4)
days. The LMC has gas phase oxygen abundance $12 + \log($O/H$)=
8.39\pm0.10$ \citep{Pagel78} while the SMC is $12 + \log($O/H$)=
7.98\pm0.10$ \citep{Peimbert76}.

The Araucaria Project \citep{Pietrzynski02} is a photometric survey of
Local and Sculptor Group galaxies and their Cepheid populations. The
primary goal is to more accurately determine the distances to these
galaxies and to characterize the dependence of various stellar
distance indicators on metallicity and age. The Araucaria Project has
observed ULP Cepheids in the following galaxies.

\textbf{NGC 55} Five ULP Cepheids were found in NGC 55
\citep{Pietrzynski06}. Observations were taken with the Optical
Gravitation Lensing Experiment (OGLE) detector on the Warsaw 1.3 m
telescope at Las Campanas Observatory. They estimate that the
calibration procedure used to transform their photometric data from
the OGLE filters to the standard system produced errors $\le0.03$
mag. Follow up observations in the IR revealed a total reddening of
$E(B-V)= 0.13$ mag \citep{Gieren08}. Their multi-wavelength PL
analysis produced a DM to NGC 55 of $26.43\pm0.04\pm0.08$ mag
(statistical and systematic errors, respectively). NGC 55 hosts ULP
Cepheids with periods of 85.1, 97.7, 112.7, 152.1, and 175.9 days. The
oxygen abundance of NGC 55 is $12 + \log($O/H$)= 8.05\pm0.10$
\citep{Tullmann03}.

\textbf{NGC 6822} One ULP Cepheid was found in NGC 6822
\citep{Pietrzynski04}. The filters and telescope used are identical to
those of NGC 55 \citep{Pietrzynski06}. Similarly, the reported
calibration error onto the standard system is $\le0.03$ mag. As in
the multi-wavelength follow up study of NGC 6822 \citep{Gieren06}, we adopt a
total reddening of $E(B-V)=0.36$ mag. The lone ULP Cepheid in NGC 6822
has a period of 123.9 days. \citet{Gieren06} calculate a DM to NGC
6822 of $23.31\pm0.02\pm0.06$ mag (statistical and systematic errors,
respectively). NGC 6822 has a similar oxygen abundance to NGC 55 of
$12 + \log($O/H$)= 8.11\pm0.10$ \citep{Peimbert05}.

\textbf{NGC 300} \citet{Gieren04} found three ULP Cepheids in NGC
300. Again, OGLE filters were used for the observations. Their
calibration onto the standard system has a reported error $\le0.03$
mag. A multi-wavelength study of NGC 300 \citep{Gieren05} determined a
reddening-free DM of $26.37\pm0.05\pm0.03$ mag (statistical and
systematic, respectively) using a total reddening of $E(B-V)=0.10$
mag. ULP Cepheids of 83.0, 89.1, and 115.8 days are observed in NGC
300. NGC 300 has a strong metallicity gradient; therefore we adopt
mean Cepheid radial distance of 4 kpc and apply the averaged gradient
of \citet{Urbaneja05} to obtain a mean oxygen abundance value of $12 +
\log($O/H$)= 8.25\pm0.22$.

\textbf{I Zw 18} \citet{Aloisi07} discovered three ULP Cepheids from
the extremely metal poor galaxy I Zw 18, though they could not confirm
one candidate. A follow-up study \citep{Fiorentino08} presents flux
weighted mean photometry but no data, so the light curves of
these objects could not be included in Figure~\ref{fig:lcs}. The ULP
Cepheids have periods of $129.8$ and $125.0$ days
(Table~\ref{tab:ceps}). After accounting for extinction $E(B-V)=0.032$
mag, \citet{Aloisi07} use the red giant branch tip to determine a DM
of $31.30\pm0.17$ mag while \citet{Fiorentino08} find a DM of
$31.35\pm0.26$ mag via pulsation models. We use the former measurement
as it is considered more reliable by the authors. We do not include
these two Cepheids in the upcoming PL determination as I Zw 18 is a
full dex more metal poor than the other galaxies in this sample ($12 +
\log($O/H$)= 7.2\pm0.10$; \citealt{Skillman93}). There is an increasing
amount of support for a metallicity dependent PL
\citep[\eg][]{Sandage08} and including these Cepheids in our PL
analysis would greatly increase the metallicity dispersion of the host
galaxies in our sample. We do, however, make use of them to examine
the ULP PL relation dependence on metallicity.

\subsection{Absolute Photometry}

\begin{figure*}
\figurenum{2}
\plotone{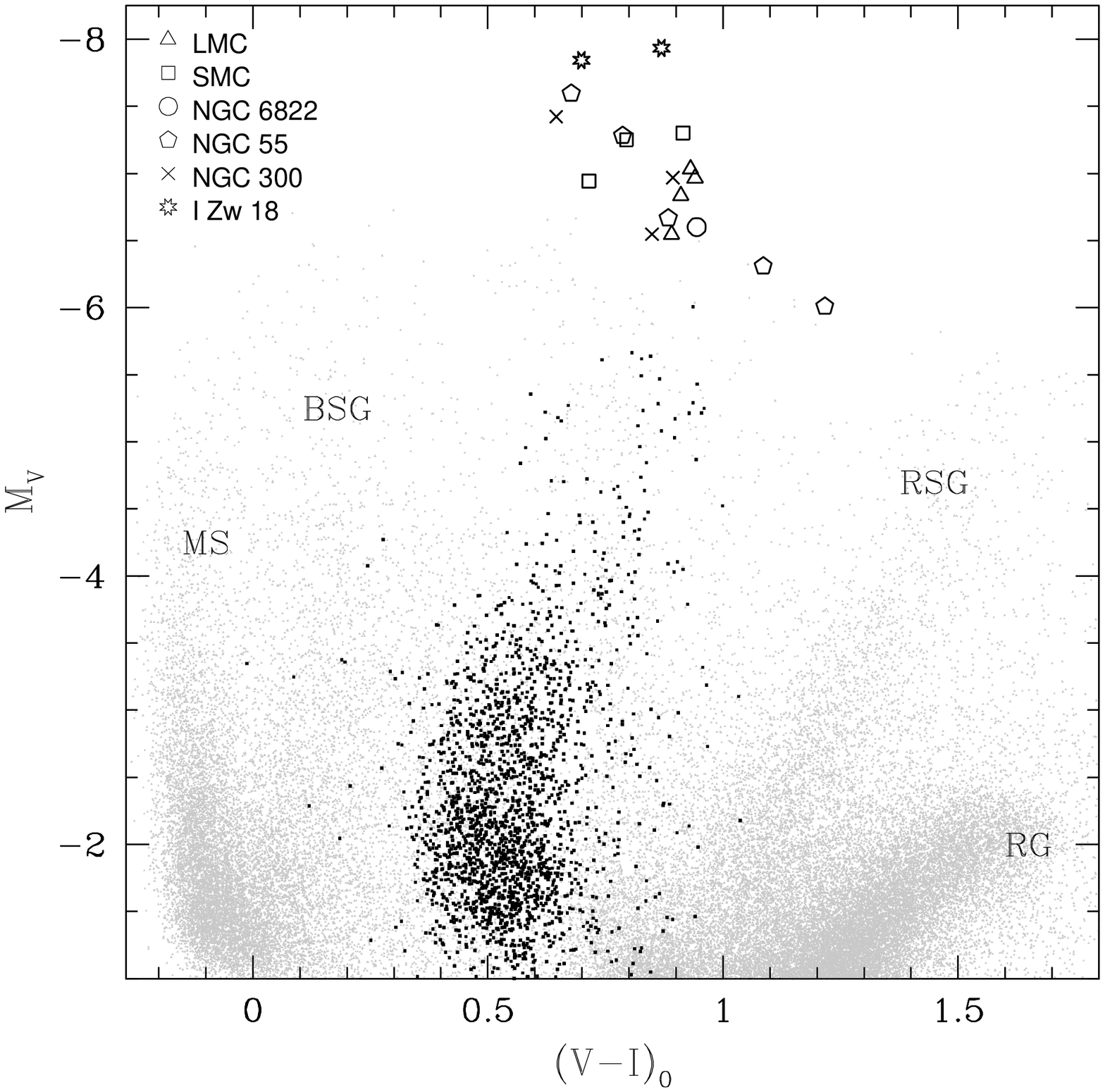}
\caption{\label{fig:cmd}M$_V$ versus $(V-I)_0$ color CMD of our ULP
sample (open symbols) with the OGLE SMC Cepheids (black dots) and OGLE
SMC stars (gray dots) for comparison. The legend denotes the host
galaxy of each ULP Cepheid. For reference we label the main sequence
(MS), blue supergiant (BSG), red supergiant (RSG), and red giant (RG)
sequences.}
\end{figure*}

The ULP Cepheid sample and its mean, flux-weighted photometry in the
standard system can be found in Table~\ref{tab:ceps}. We assume that
the photometric error associated with each ULP Cepheid is negligible
when compared to the intrinsic scatter of the PL relation. We
transform these measurements to absolute magnitudes via:
\begin{equation}
M_i=m_i-DM-A_i,\ i=I,V \label{eq:abs_mag}
\end{equation}
where $M_i$ is the absolute magnitude in either the $V$ or $I$, $m_i$
is the apparent magnitude; $DM$ is the reddening free distance
modulus; and $A_i$ is the extinction in the $V$ or $I$. We use the
extinction law $A_V=3.24\,E(B-V)$ and $A_I=1.96\,E(B-V)$
\citep{Schlegel98}. We define the Wesenheit magnitudes as:
W$_I=I-1.55(V-I)$ \citep[\eg,][]{Udalski99}.

The color-magnitude diagram (CMD) highlights several important
characteristics of the ULP Cepheid data set
(Figure~\ref{fig:cmd}). ULP Cepheids are the luminous counterparts to
shorter period Cepheids in color magnitude space. Our sample clearly
populates the luminous region of the instability strip. Future Cepheid
studies can use ULP Cepheids to push Cepheid distance measurements
well beyond the current $\sim30$ Mpc limit as our sample's median
absolute magnitude is M$_I$(M$_V)=-7.86(-6.97)$ (see
Section~\ref{sec:conclusion}). The intrinsic brightness of ULP
Cepheids makes them ideal candidates for distance indicators to
galaxies where classical Cepheids cannot be observed.

\section{Distance Measurements with ULP Cepheids} \label{sec:DM}

The Cepheids in our sample have been ignored as distance indicators as
they do not follow to the standard period-luminosity relationship
\citep[\eg,][]{Freedman92}. In this section, we examine the
characteristics of our ULP Cepheid sample in the period-luminosity
plane and explore their viability as a distance indicator. We
determine PL relations of our sample in the $V$, $I$, and W$_I$ in
Section~\ref{sec:PL} while the metallicity dependence of our results
is presented in Section~\ref{sec:metal}.

\subsection{Period Luminosity Relations} \label{sec:PL}

\begin{figure*}
\figurenum{3}
\plotone{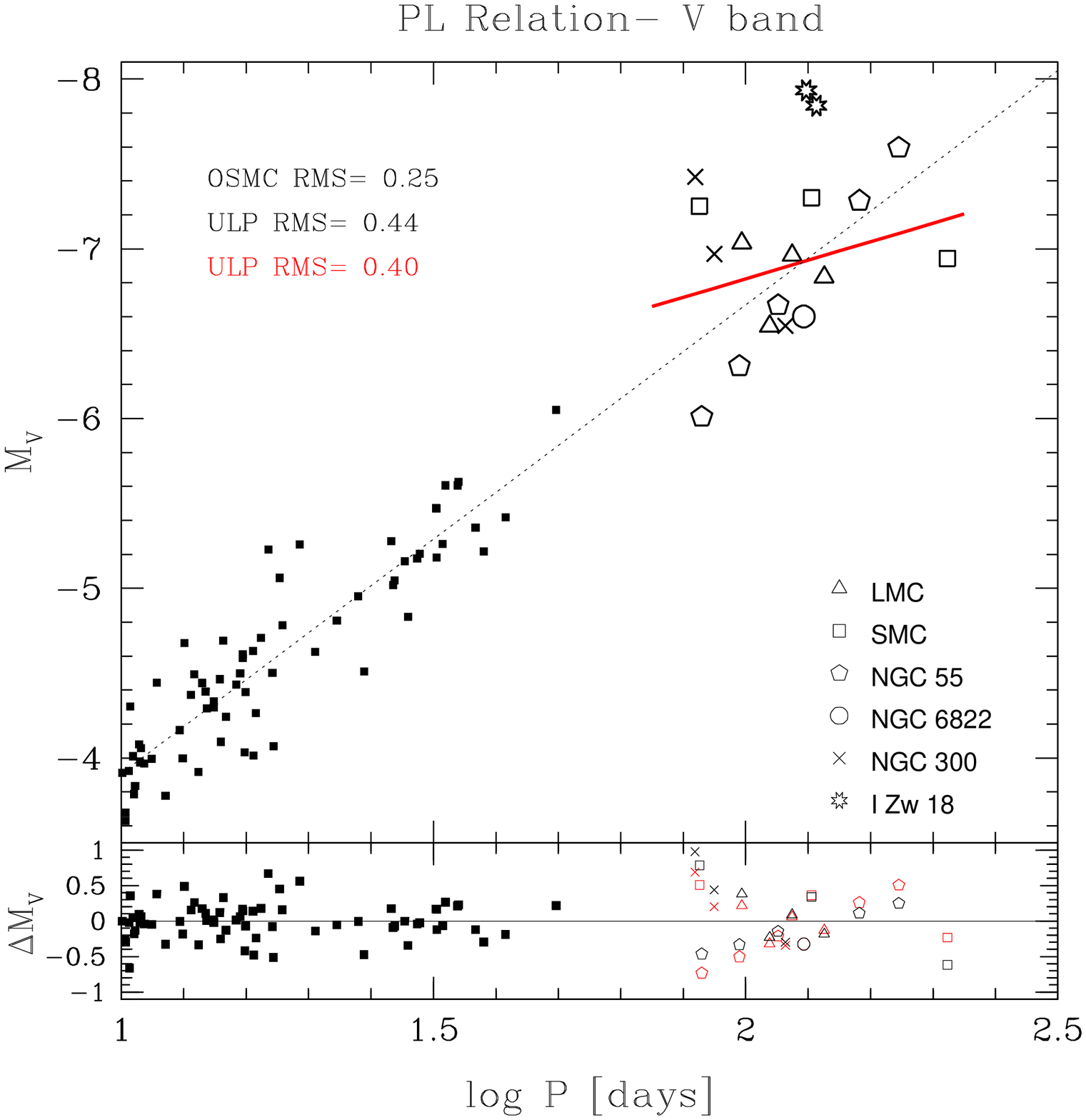}
\caption{\label{fig:PL_V}The $V$ Period Luminosity relationship for
the OGLE SMC Cepheids (black dots) and ULP Cepheids (open
symbols). The dashed line is the PL relation adopted from
\citet{Udalski99}, with a slope of $-2.76$. The least square fit of
the ULP Cepheid subsample yields a flatter slope of $-1.09\pm0.94$
(red line) and the RMS is $0.40\;$mag. The residuals to the \pls\ fit
are shown in the bottom panel (black, open squares). Residuals to the
\plu\ are given for the ULP Cepheid sample (red symbols).}
\end{figure*}

\begin{figure*}
\figurenum{4}
\plotone{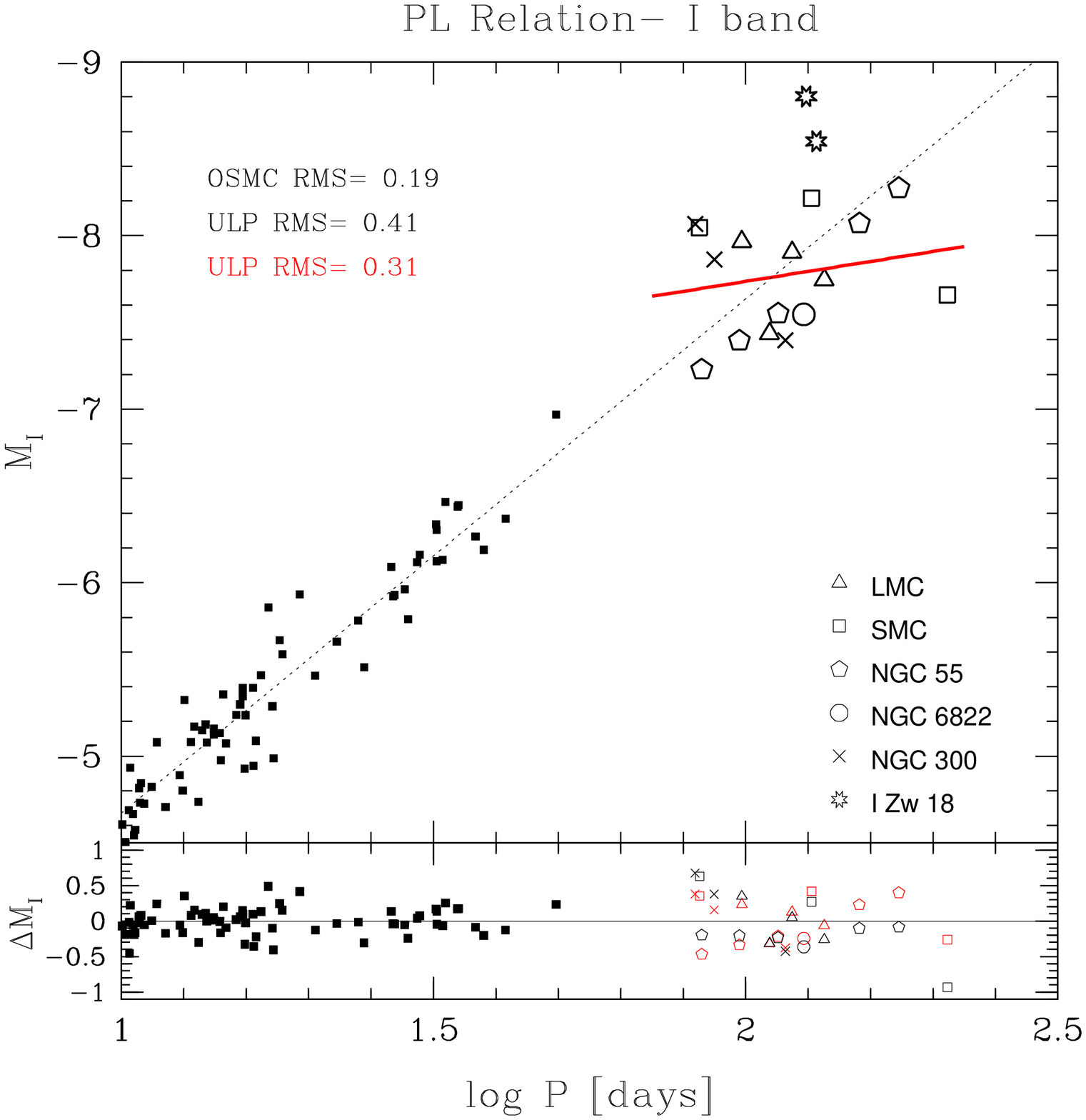}
\caption{\label{fig:PL_I}The $I$ Period Luminosity relationship for
the OGLE SMC Cepheids (black dots) and ULP Cepheids (open
symbols). The dashed line has a slope of $-2.96$ and is the PL
relation adopted from \citet{Udalski99}. The least square fit to the
ULP Cepheid subsample produces a flatter slope of $-0.57\pm0.73$ (red
line) and the RMS is $0.31$ mag. The residuals to the \pls\ fit are
shown in the bottom panel (black, open squares). Residuals to the
\plu\ are given for the ULP Cepheid sample (red symbols).}
\end{figure*}

\begin{figure*}
\plotone{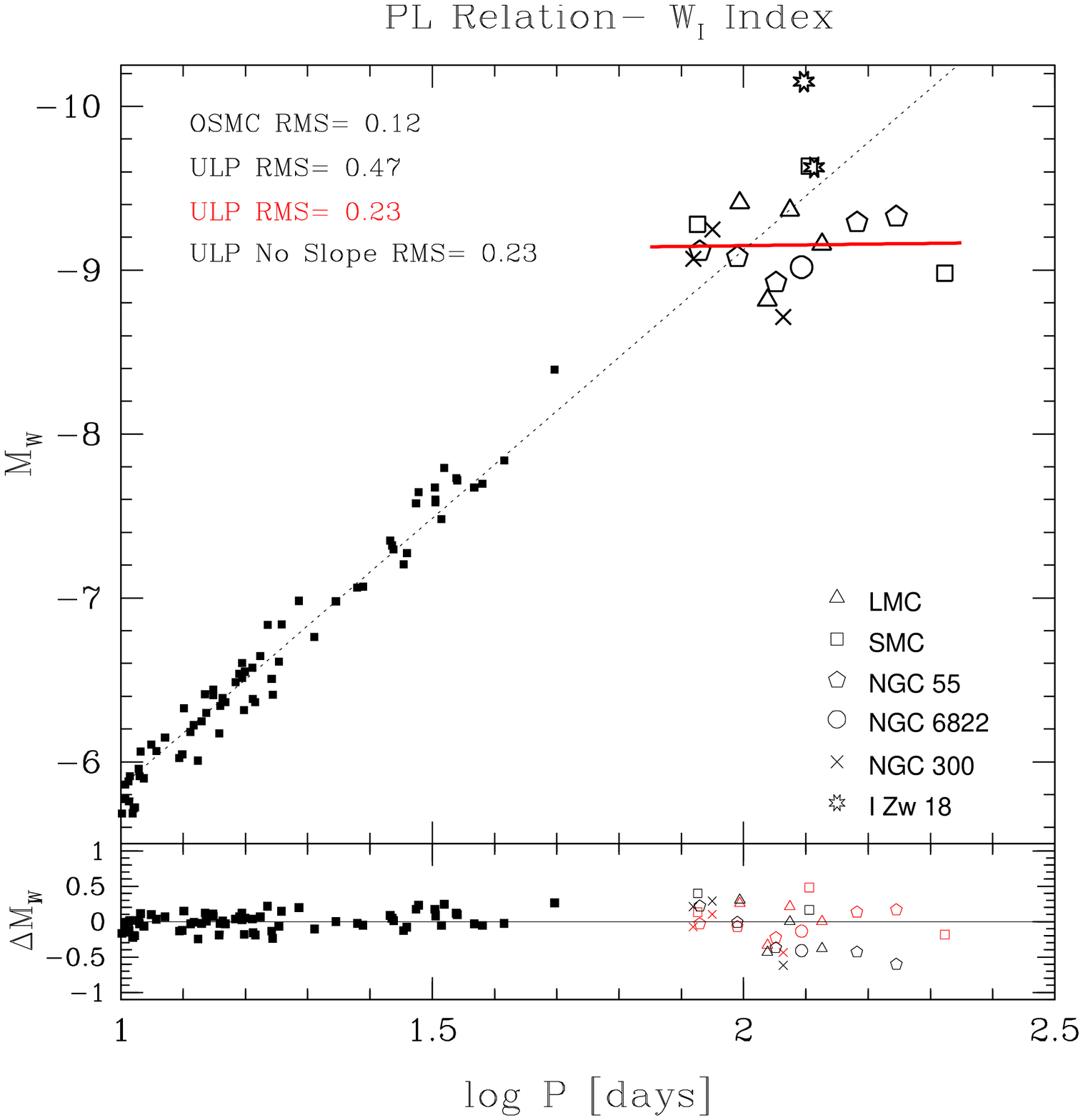}
\figurenum{5}
\caption{\label{fig:PL}The absolute Wesenheit magnitude Period
Luminosity relationship for the OGLE SMC Cepheids (black dots) and ULP
Cepheids (open symbols). The dashed line is the PL relation adopted
from \citet{Udalski99} and has a slope of $-3.28$. The least square
fit to the ULP Cepheid subsample produces a flat slope of
$-0.05\pm0.54$ (red line) with a ULP Cepheid RMS $= 0.23$ mag.  If we
assume that \plu\ has zero slope (intercept of $-9.15$), the RMS stays
at 0.23 mag. The residuals to the \pls\ fit are shown in the bottom
panel (black, open squares). Residuals to the \plu\ are given for the
ULP Cepheid sample (red symbols).}
\end{figure*}

Using the data in Table~\ref{tab:ceps}, we construct PL diagrams in
$V$, $I$, and W$_I$ (Figures~\ref{fig:PL_V},~\ref{fig:PL_I},
~\ref{fig:PL}, respectively). In each case, ULP Cepheids are compared
to fundamental mode SMC Cepheids (hereafter, this control sample will
be referred to as OGLE Cepheids). The \osmc\ sample contains over
$1100$ fundamental mode Cepheids with periods ranging from $0.5$ to
$\sim50$ days; however, we only plot the 70 Cepheids with periods of
greater than 10 days. To quantify our comparison, we perform a linear
fit on both samples in each PL diagram: the slopes of the \osmc\ PL
relations (hereafter PL$_{SMC}$; dotted lines in the figures) are
adopted from \citet{Udalski99} while the intercepts are chosen to
minimize chi-square. We employ linear least square fitting of the ULP
Cepheid sample to identify their PL relation (hereafter
PL$_{ULP}$). We omit errors in distance moduli and extinction in this
demonstration as they are small when compared to the overall scatter
of the sample. The parameters of these fits and the RMS of each data
set in $V$, $I$, and W$_I$ are listed in Table~\ref{tab:fits}. Despite
our sample ranging in period from 83 to 210 days, ULP Cepheids occupy
a small region of luminosity space. We note that the ULP Cepheids from
I Zw 18 are not included in this analysis for reasons outlined in
Section~\ref{sec:sample}.

The $V$ PL diagram is Figure~\ref{fig:PL_V}.  The \pls\ fit has a
slope of $-2.76$ and the RMS of the \osmc\ sample about this fit is
$0.25$ mag. The ULP Cepheid sample has $76\%$ more scatter (RMS=$0.44$
mag) about the \pls\ fit. This discrepancy in scatter has led to the
standard practice of removing ULP Cepheids from PL relation studies
\citep[\eg,][]{Freedman85} and the significant increase in RMS
suggests that the ULP Cepheids do not conform to the classical PL
relation. If we determine the PL relation of ULP Cepheids alone we
find \plu\ has a slope ($-1.09\pm0.94$) that is flatter than \pls\
(though the slopes are within $2\sigma$ of each other, see
Table~\ref{tab:fits}). The RMS of our sample to \plu\ is $0.40$ mag;
only marginally better than the ULP Cepheid scatter about the
established \pls\ relation. In $V$, the ULP Cepheid sample does not
follow a statistically distinct and unique PL.

The longer the wavelength the less reddening is a concern. The
accuracy of distance measurement with Cepheids increases going from
$V$ to $I$ (PL diagram in Figure~\ref{fig:PL_I}). \pls\ has a slope of
$-2.96$ and the \osmc\ sample's RMS is $0.19$. The ULP Cepheid scatter
about this fit is $116\%$ larger ($0.41$ mag). The RMS of the ULP
Cepheids is reduced to $0.31$ mag when using the \plu\ fit
(slope$=-0.57\pm0.73$). \plu\ is approximately five times as flat as
\pls\ and the two slopes are distinct at the $3\sigma$ level. While
ULP Cepheids show the same general trends with regards to period,
luminosity, and color as normal Cepheids, significant statistical
differences between the two populations are apparent in the $I$-band
PL relation.

To further reduce the uncertainty associated with reddening in our PL
analysis, we repeat the procedure using the reddening-free Wesenheit
Index (W$_I$) introduced in \citet{Madore91}. The W$_I$ PL diagram
illuminates the advantages of this reddening-free index for Cepheid
distance measurements (Figure~\ref{fig:PL}). The \osmc\ sample has a
very tight relation between period and luminosity; with small scatter
about the \pls\ fit (slope of $-3.28$; RMS of only $0.12$ mag).  The
ULP Cepheid RMS about \pls\ is $0.47$ mag. The $4$-fold increase in
scatter implies the ULP Cepheid and the \osmc\ samples do not conform
to the same PL relation. The \plu\ fit slope is flatter than its \pls\
counterpart at the $6\sigma$ level ($-0.05\pm0.54$ vs. $-3.28$) and
the ULP fit slope is consistent with zero slope. The scatter of the
ULP Cepheids is only $0.23$ mag about the \plu\ relation. This scatter
is still $92\%$ more than that of the \osmc\ sample; however, the ULP
sample is relatively small and heterogeneous. We note that the scatter
of ULP Cepheids about the \plu\ fit is smaller than that of the \osmc\
sample about the nominal PL relation in $V$ and on par with the same
in $I$.

Several trends in PL space are apparent as one examines the ULP
Cepheid sample in $V$, then $I$, and finally in the W$_I$ index. As
reddening is reduced, the \plu\ parameters are increasingly different
from those of \pls. The \plu\ fit is more shallow as one moves from
$V$ to W$_I$. In the W$_I$, the \plu\ relation reveals that ULP
Cepheid luminosity becomes statistically independent of period. In
essence, ULP Cepheids behave as bright, standard candles in the
reddening-free index.

\subsection{Metallicity Dependence}\label{sec:metal}

\begin{figure*}
\figurenum{6}
\plotone{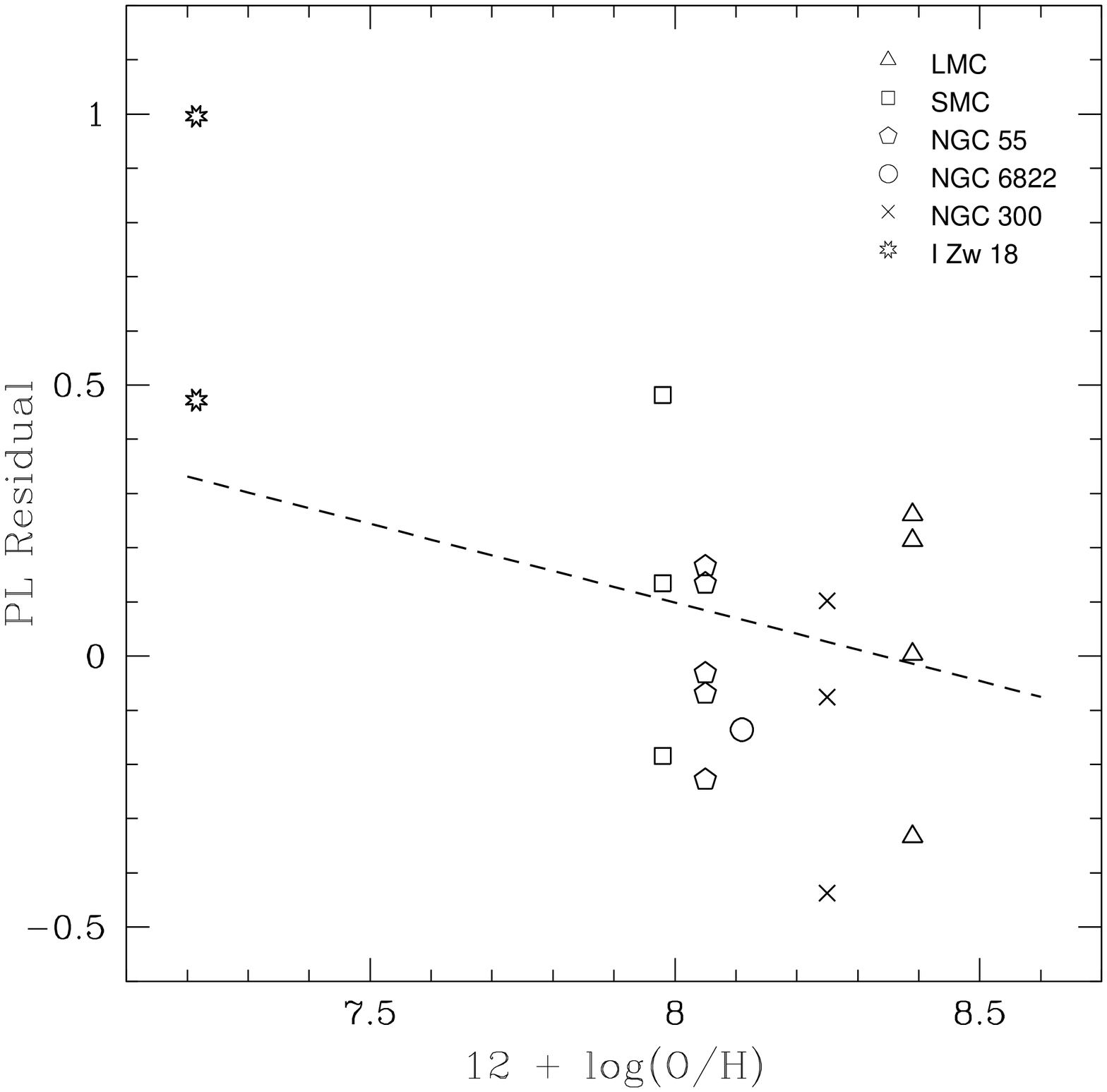}
\caption{\label{fig:res_OH} Residual to the Wesenheit index PL
relation (See Table~\ref{tab:fits}, PL$_{ULP,W_I}$) versus
metallicity. The open symbols are ULP Cepheids from the sample. The
dotted line corresponds to the luminosity correction of $-0.29\pm0.10$
mag dex$^{-1}$ determined by \citet{Macri06} and normalized to our
sample. The residuals of the sample, minus the two I Zw 18 Cepheids,
are consistent with the \citet{Macri06} result.}
\end{figure*}

An uncertainty of the Cepheid PL and its derived distance measurement
is its sensitivity to the metallicity of the stars
\citep[\eg,][]{Freedman90, Kennicutt98}. Our sample of ULP Cepheids
contains six different host galaxies that span a range of $\sim1.2$
dex in $12 + \log($O/H$)$ from 7.22 to 8.39, providing an opportunity
to investigate the dependence of the ULP Cepheid PL on metallicity. We
plot the residual of each ULP Cepheid to the PL$_{ULP,W_I}$ fit listed
in Table~\ref{tab:fits} as a function of metallicity
(Figure~\ref{fig:res_OH}). In other studies, linear fits of PL
residuals have determined a correction factor, $\gamma$, between $0.0$
and $-0.4$ mag dex$^{-1}$ \citep[see Figure 1
of][]{Romaniello08}. Recently, \citet{Macri06} used the metallicity
gradient in NGC 4258 to determine $\gamma = -0.29\pm0.09\pm0.05$ mag
dex$^{-1}$ (random and systematic errors, respectively). We overlay
this relation (normalized to our data set) in Figure~\ref{fig:res_OH}
for reference. If the I Zw 18 Cepheids are confirmed, it suggests a
stronger correlation between PL offset and metallicity than is evident
in lower period Cepheids \citep[\eg,][]{Kochanek97, Kennicutt98}.
However, we note that we do not take into account any reddening or DM
errors in this analysis. As such, we do not claim a specific
metallicity dependence for ULP Cepheids. We simply demonstrate that
the residuals to the \plu\ fit are broadly consistent with the range
of values presented in the literature to date.

\tabletypesize{\tiny}
\begin{deluxetable}{llllll}[ht]
\tablecolumns{6}
\tablewidth{3.5in}
\tablecaption{Least Square Fit Values: $y = b + a*x$}
\tablehead{
\colhead{Relation} &
\colhead{Subset} &
\colhead{Shorthand} &
\colhead{Intercept (b)} &
\colhead{Slope (a)} &
\colhead{RMS}\\
\colhead{(1)} &
\colhead{(2)} &
\colhead{(3)} &
\colhead{(4)} &
\colhead{(5)} &
\colhead{(6)} \\
}
\startdata
Period-Luminosity: $V$ & SMC Cepheids & PL$_{SMC,V}$ & $-1.15$ & $-2.76$ & $0.25$ \\
Period-Luminosity: $V$ & ULPs & PL$_{ULP,V}$ & $-4.64\pm1.95$ & $-1.09\pm0.94$ & $0.40$ \\
Period-Luminosity: $I$ & SMC Cepheids & PL$_{SMC,I}$ & $-1.71$ & $-2.96$ & $0.19$ \\
Period-Luminosity: $I$ & ULPs & PL$_{ULP,I}$ & $-6.58\pm1.50$ & $-0.57\pm0.73$ & $0.31$ \\
Period-Luminosity: $W_I$ index & SMC Cepheids & PL$_{SMC,W_I}$ & $-2.57$ & $-3.28$ & $0.12$ \\
Period-Luminosity: $W_I$ index & ULPs & PL$_{ULP,W_I}$ & $-9.06\pm1.12$ & $-0.05\pm0.54$ & $0.23$ \\
\enddata													        

\tablecomments{\label{tab:fits} The fit values of the PL relationships
in $V$, $I$, and W$_I$. For each photometric system, we calculate the
PL relation of classical SMC (PL$_{SMC}$) and ULP (PL$_{ULP}$)
Cepheids.}
\end{deluxetable}

At this time we do not apply a metallicity correction to our ULP
Cepheid PL relations. Precise gas phase oxygen abundance measurements
are difficult to obtain \citep[for a review see][]{Bresolin06} and we
adopt a minimum metallicity error of $0.1$ dex. The ULP Cepheid sample
must grow in size and the PL analysis must be more detailed to
determine if I Zw 18 is an exception and to characterize the
functional form of the ULP Cepheid PL sensitivity to metallicity.

\section{Using ULP Cepheids to Probe the Evolutionary Models of Massive Stars} \label{sec:massloss}

\begin{figure*}[p]
\figurenum{7}
\plotone{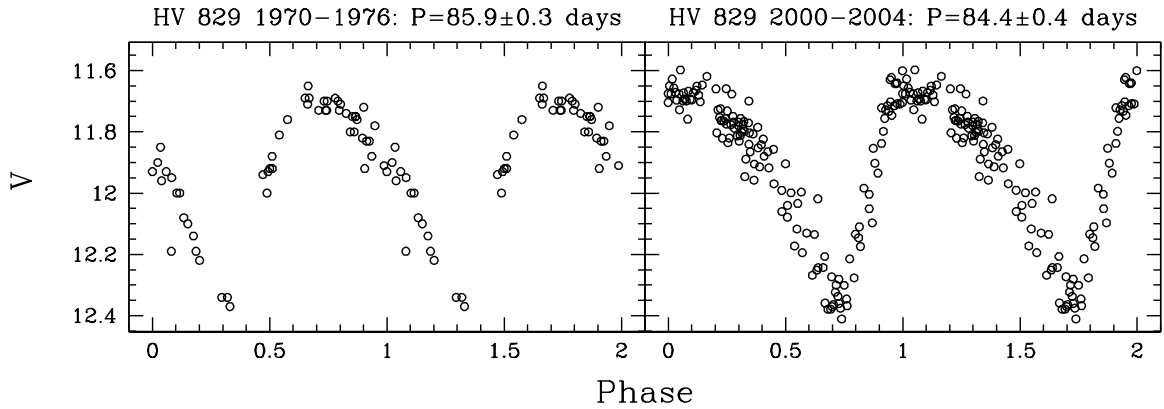}
\caption{\label{fig:dpdt} $V$-band photometry of SMC HV829 taken from
1970-1976 (left panel) and 2000-2004 (right panel; see text for data
references). Each set of data was phased with the period shown at the
top of each pane. The pulsation period has clearly decreased, from
85.9 days to 84.4 days over approximately 30 years.}
\end{figure*}

Most Cepheid variables cross the instability strip three times
\citep[\eg,][]{Bono00,Pietrukowicz01}. One can determine which
crossing a Cepheid is undergoing by measuring its period change,
$dP/dt$. The first and third crossings are associated with positive
$dP/dt$ while Cepheids exhibit a decreasing period on their second
crossing. We investigated the light curves of our sample for signs of
period changes by comparing photometry taken over the last $30$
years. One ULP Cepheid, HV829, exhibits a negative period change of
about 1.5 days. For this Cepheid, we compiled photometric data taken
during 1970-1976 from \citet{Madore75} and \citet{vanGenderen83} and
2000-2004 data from the ASAS catalog (See Figure~\ref{fig:dpdt}). This
result is confirmed by a measured period of $87.63$ days in
\citet{Payne66} and 85.2 days by \citet{Moffett98}, firmly
establishing HV829 as a Cepheid on its second crossing. No other
Cepheid in our sample exhibited a measurable period change.

ULP Cepheids occupy a mass range that is ideal to probe high mass
evolutionary models along the instability strip. We plot the CMD of
our ULP Cepheid sample and overlay the evolutionary tracks of
\citet{Lejeune01} (Figure~\ref{fig:mass}). The location of ULP
Cepheids in the color magnitude diagram suggest masses between $13$
and $20 M_{\odot}$, depending on the assumed metallicity (here, we
choose SMC metallicity). Our sample clearly probes a mass range
unexplored by current Cepheid studies. Note the evolutionary tracks at
$15$ and $20 M_{\odot}$ only cross the instability strip once,
regardless of assumed mass-loss rate. However, we have shown and
confirmed that HV829 is undergoing its \emph{second} crossing; a
result at odds with the models. We note that the evolutionary model
represents some ``mean'' behavior based on assumptions of stellar
chemical composition, overshooting, and other
parameters. Nevertheless, our result is one of the few observational
constraints on high mass stellar evolutionary models. HV829 may
indicate that massive stars behave differently than expected. Future
stellar evolutionary models in this mass regime should take this into
account.

\section{Discussion}\label{sec:conclusion}

We have presented, for the first time, a collection of Ultra Long
Period ($P>80\;$days) Cepheids from the literature and demonstrated
their viability as a distance indicator. In the past, ULP Cepheids
have been ignored as distance indicators, and in fact many stellar
variability searches did not extend their cadence and search strategy
to allow for their discovery.

In $V$, $I$, and especially reddening-free W$_I$ magnitude, ULP
Cepheids have a relatively flat PL relation compared to the respective
relations for classical Cepheids \citep{Udalski99}.  The dispersion in
both the classical and ULP Cepheid populations about their respective
PL relationships becomes smaller as one moves from $V$ to $I$ to
W$_I$, and the discrepancy between the slopes increases. Most notably,
the slope of the ULP Cepheid W$_I$ PL relation is significantly
shallower than the standard SMC PL relation (slope is $-0.05$
vs. $-3.28$). The reddening-free Wesenheit index produces the tightest
ULP PL relation, with RMS residual of only $0.23$ mag (See
Figure~\ref{fig:PL}). Other papers \citep[\eg][]{Kanbur07} have found
non-linearites in the PL relationship at lower periods. However, the
change in slope seen here is much more dramatic and it is unlikely
that it shares a physical connection with the PL changes at lower
periods.

Our ULP PL scatter in W$_I$ is already less than that of the initial
peak brightness vs. absolute magnitude relation for Type Ia Supernova
($\sim 0.3$ mag; \cite{Phillips93}).  A huge amount of effort has
been necessary to increase the number of observed Type Ia SNe,
refine the calibration technique \citep[\eg,][]{Prieto06}, and to obtain
the low $0.15$-$0.20$ mag scatter in the relation seen today
\citep[\eg,][]{Jha07}. We expect future observational and theoretical
studies of ULP Cepheids will further decrease the already fairly small
scatter found in this first analysis.

We strove to find all the known ULP Cepheids in the literature, but
our sample is likely not a complete census of ULP Cepheids. We note
that the Araucaria Project found ULP Cepheids in three of the five
galaxies they observed at the time of this analysis, so the ULP
Cepheid sample should grow at a reasonable rate as Cepheid studies are
extended to more galaxies and previous data sets are reanalyzed with
longer period searches. There is also evidence ULP Cepheids exist in a
broad range of metallicities, as preliminary data analysis of a M81
variability survey with the Large Binocular Telescope has already
discovered at least one ULP Cepheid (Kochanek, private communication).

In addition to following a fairly tight PL relation, the ULP Cepheids
are also very luminous, with a median absolute magnitude in $I$($V$)
for our sample of $-7.86(-6.97)$. Using WFPC3, {\em HST}\/ can obtain
$10\%$ photometry at $V=28$ or $I=27$ (DM $= 35$ --- distance of
$100\;$Mpc for the median ULP Cepheid) in about 10 orbits, while only
two orbits are needed to reach a DM of $34$. As one would only need a
few orbits per epoch, one could detect the median ULP Cepheid (with a
period of $\sim121$ days) at 100 Mpc and the brightest ULP Cepheids
at $\sim150$ Mpc in a reasonable amount of time. Since the luminosity
of an ULP Cepheid is a weak function of its period, relatively
accurate distances would not require as precise period measurements as
is needed for classical Cepheids. We encourage future variability
proposals to search for Cepheids with periods greater than 100 days.

We note two concerns in using ULP Cepheids for distance
measurements. It is obvious from our sample size that ULP Cepheids are
far less common than classical Cepheids. The relatively small ULP
Cepheid population will make precision distance measurements less
practical. As the sample size grows and the ULP Cepheid PL relations
become established, a single ULP Cepheid observation may yield
distance measurements accurate to $10-20\%$. Blending is always a
concern in Cepheid studies. Blending can compromise Classical Cepheid
measurements as stars of comparable brightness are likely to lie
within a single PSF \citep{Mochejska00}, especially at larger
distances. We expect blending to be less of a problem for our ULP
Cepheids even at large distances simply because they are so bright and
there are very few stars of comparable brightness in a given
galaxy. The effect of blending on ULP Cepheid observations will need
to be investigated further as the sample is increased.

\begin{figure*}[p]
\epsscale{0.80}
\figurenum{8}
\plotone{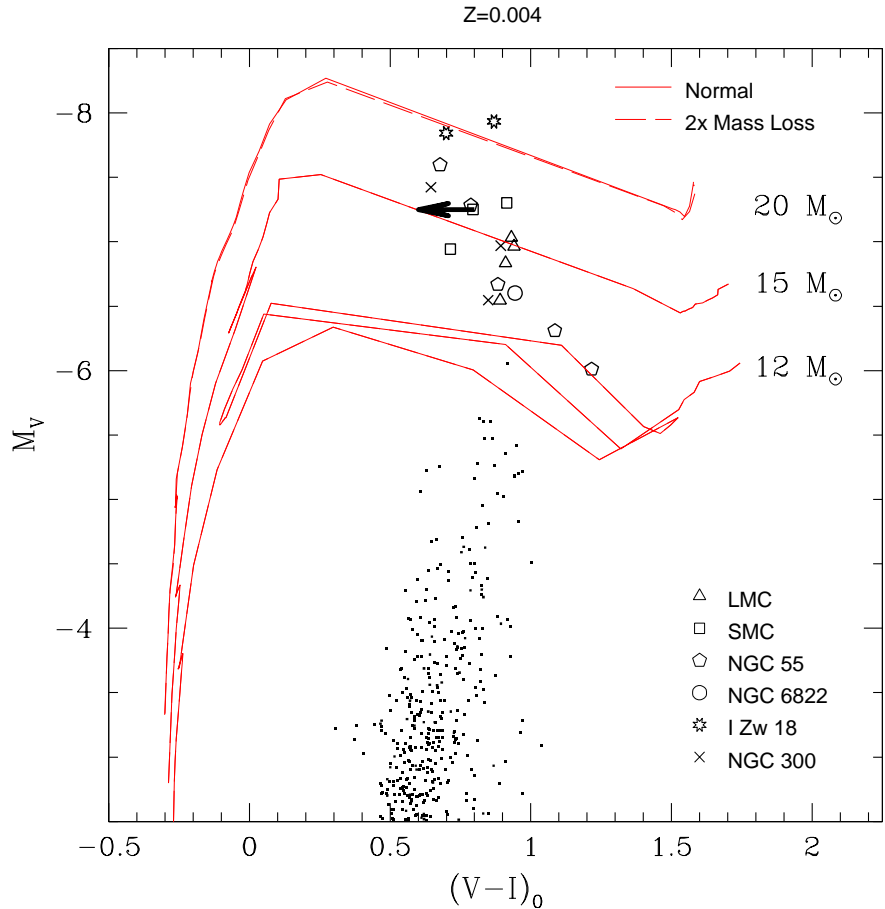}
\caption{\label{fig:mass} CMD of OGLE SMC Cepheids (small blacks dots)
and ULP Cepheids (large open symbols) overlaid with 12 M$_{\odot}$, 15
M$_{\odot}$, and 20 M$_{\odot}$ evolutionary models of
\citet{Lejeune01}. The metallicity chosen for the models is similar to
the SMC. For each mass, models incorporating ``normal'' (solid line)
mass loss and ``double'' (dashed line) mass loss are plotted.  The two
mass loss models produce essentially the same evolutionary track for
each mass plotted. The large arrow denotes that the pulsation period
of HV829 is becoming smaller; therefore the star is undergoing its
second crossing of the instability strip.}
\end{figure*}

The data set and analysis herein provides meaningful constraints on
theoretical work on Cepheids in this mass and period range. We have
shown that ULP Cepheids show strong evidence for a different, flatter
PL relation than their lower period cousins. Several papers have
modeled Cepheid pulsations and mapped these to theoretical PL relations
\citep[\eg,][]{Bono02}. However, this work has not been reliably
extended to the period range of our sample. Pulsation models in this
period range would also help determine the intrinsic scatter to the
PL$_{ULP}$ relation.

Our current sample is not large enough to constrain the sensitivity of
the ULP Cepheid PL to metallicity. The PL residuals as a function of
metallicity are consistent with the results for shorter period
Cepheids \citep[$\gamma= -0.29 \pm0.09\pm 0.05$ mag
dex$^{-1}$;][]{Macri06}. Note that the median $12 + \log($O/H$)$ value
for a galaxy in our sample is about $0.5\;$dex lower than the
corresponding value for the {\em HST}\/ Key Project Cepheid hosts.

Period changes of ULP Cepheids have powerful implications for stellar
evolutionary models in this mass regime. We examined photometry from
two epochs separated by thirty years to check for evidence of period
changes in seven Magellanic Cloud ULP Cepheids. Only one, SMC HV829,
showed a significant period change--- from $85.9\pm0.3$ to $84.4\pm0.4$
days. The negative period derivative indicates that
the Cepheid is undergoing its second crossing. Assuming the
metallicity of the SMC, many current evolutionary models do not
predict second crossings for Cepheids in this mass range (See
Figure~\ref{fig:mass}). Future models should incorporate this
observational result as it should place limits on several key input
parameters. In addition to their potential as distance indicators, ULP
Cepheids provide observational constraints on high mass stellar
evolution models.

To summarize, ULP Cepheids, while often dismissed in the past, are
potentially a powerful distance indicator, probe of PL metallicity
sensitivity, and also a probe of massive star evolutionary
models. ULP Cepheids could provide the first direct stellar distance
measurements to galaxies in the 50-150 Mpc range, extending the
cosmological distance ladder well into the Hubble flow.

\acknowledgements{ We would like to thank Chris Kochanek for his
thoughtful comments on the manuscript. We thank the referee for
improving this work with helpful comments. This work was supported by
NSF grant AST-0707982.}

\bibliographystyle{apj}

\begin{thebibliography}{55}
\expandafter\ifx\csname natexlab\endcsname\relax\def\natexlab#1{#1}\fi

\bibitem[{{Aloisi} {et~al.}(2007){Aloisi}, {Clementini}, {Tosi}, {Annibali},
  {Contreras}, {Fiorentino}, {Mack}, {Marconi}, {Musella}, {Saha}, {Sirianni},
  \& {van der Marel}}]{Aloisi07}
{Aloisi}, A., {Clementini}, G., {Tosi}, M., {Annibali}, F., {Contreras}, R.,
  {Fiorentino}, G., {Mack}, J., {Marconi}, M., {Musella}, I., {Saha}, A.,
  {Sirianni}, M., \& {van der Marel}, R.~P. 2007, \apjl, 667, L151

\bibitem[{{Baade}(1956)}]{Baade1956}
{Baade}, W. 1956, \pasp, 68, 5

\bibitem[{{Bessel}(1839)}]{Bessel1839}
{Bessel}, F.~W. 1839, Astronomische Nachrichten, 16, 65

\bibitem[{{Bono} {et~al.}(2000){Bono}, {Caputo}, {Cassisi}, {Marconi},
  {Piersanti}, \& {Tornamb{\`e}}}]{Bono00}
{Bono}, G., {Caputo}, F., {Cassisi}, S., {Marconi}, M., {Piersanti}, L., \&
  {Tornamb{\`e}}, A. 2000, \apj, 543, 955

\bibitem[{{Bono} {et~al.}(2002){Bono}, {Groenewegen}, {Marconi}, \&
  {Caputo}}]{Bono02}
{Bono}, G., {Groenewegen}, M.~A.~T., {Marconi}, M., \& {Caputo}, F. 2002,
  \apjl, 574, L33

\bibitem[{{Bresolin}(2006)}]{Bresolin06}
{Bresolin}, F. 2006, ArXiv Astrophysics e-prints, astroph/0608410

\bibitem[{{Chiosi} {et~al.}(1993){Chiosi}, {Wood}, \& {Capitanio}}]{Chiosi93}
{Chiosi}, C., {Wood}, P.~R., \& {Capitanio}, N. 1993, \apjs, 86, 541

\bibitem[{{Fiorentino} {et~al.}(2008){Fiorentino}, {Clementini}, {Contreras},
  {Marconi}, {Musella}, {Tosi}, {Aloisi}, {Annibali}, \& {Saha}}]{Fiorentino08}
{Fiorentino}, G., {Clementini}, G., {Contreras}, R., {Marconi}, M., {Musella},
  I., {Tosi}, M., {Aloisi}, A., {Annibali}, F., \& {Saha}, A. 2008, Memorie
  della Societa Astronomica Italiana, 79, 461

\bibitem[{{Freedman} {et~al.}(1985){Freedman}, {Grieve}, \&
  {Madore}}]{Freedman85}
{Freedman}, W.~L., {Grieve}, G.~R., \& {Madore}, B.~F. 1985, \apjs, 59, 311

\bibitem[{{Freedman} \& {Madore}(1990)}]{Freedman90}
{Freedman}, W.~L. \& {Madore}, B.~F. 1990, \apj, 365, 186

\bibitem[{{Freedman} {et~al.}(2001){Freedman}, {Madore}, {Gibson}, {Ferrarese},
  {Kelson}, {Sakai}, {Mould}, {Kennicutt}, {Ford}, {Graham}, {Huchra},
  {Hughes}, {Illingworth}, {Macri}, \& {Stetson}}]{Freedman01}
{Freedman}, W.~L., {Madore}, B.~F., {Gibson}, B.~K., {Ferrarese}, L., {Kelson},
  D.~D., {Sakai}, S., {Mould}, J.~R., {Kennicutt}, Jr., R.~C., {Ford}, H.~C.,
  {Graham}, J.~A., {Huchra}, J.~P., {Hughes}, S.~M.~G., {Illingworth}, G.~D.,
  {Macri}, L.~M., \& {Stetson}, P.~B. 2001, \apj, 553, 47

\bibitem[{{Freedman} {et~al.}(1992){Freedman}, {Madore}, {Hawley}, {Horowitz},
  {Mould}, {Navarrete}, \& {Sallmen}}]{Freedman92}
{Freedman}, W.~L., {Madore}, B.~F., {Hawley}, S.~L., {Horowitz}, I.~K.,
  {Mould}, J., {Navarrete}, M., \& {Sallmen}, S. 1992, \apj, 396, 80

\bibitem[{{Gieren} {et~al.}(2006){Gieren}, {Pietrzy{\'n}ski}, {Nalewajko},
  {Soszy{\'n}ski}, {Bresolin}, {Kudritzki}, {Minniti}, \&
  {Romanowsky}}]{Gieren06}
{Gieren}, W., {Pietrzy{\'n}ski}, G., {Nalewajko}, K., {Soszy{\'n}ski}, I.,
  {Bresolin}, F., {Kudritzki}, R.-P., {Minniti}, D., \& {Romanowsky}, A. 2006,
  \apj, 647, 1056

\bibitem[{{Gieren} {et~al.}(2005){Gieren}, {Pietrzy{\'n}ski}, {Soszy{\'n}ski},
  {Bresolin}, {Kudritzki}, {Minniti}, \& {Storm}}]{Gieren05}
{Gieren}, W., {Pietrzy{\'n}ski}, G., {Soszy{\'n}ski}, I., {Bresolin}, F.,
  {Kudritzki}, R.-P., {Minniti}, D., \& {Storm}, J. 2005, \apj, 628, 695

\bibitem[{{Gieren} {et~al.}(2008){Gieren}, {Pietrzy{\'n}ski}, {Soszy{\'n}ski},
  {Bresolin}, {Kudritzki}, {Storm}, \& {Minniti}}]{Gieren08}
{Gieren}, W., {Pietrzy{\'n}ski}, G., {Soszy{\'n}ski}, I., {Bresolin}, F.,
  {Kudritzki}, R.-P., {Storm}, J., \& {Minniti}, D. 2008, \apj, 672, 266

\bibitem[{{Gieren} {et~al.}(2004){Gieren}, {Pietrzy{\'n}ski}, {Walker},
  {Bresolin}, {Minniti}, {Kudritzki}, {Udalski}, {Soszy{\'n}ski}, {Fouqu{\'e}},
  {Storm}, \& {Bono}}]{Gieren04}
{Gieren}, W., {Pietrzy{\'n}ski}, G., {Walker}, A., {Bresolin}, F., {Minniti},
  D., {Kudritzki}, R.-P., {Udalski}, A., {Soszy{\'n}ski}, I., {Fouqu{\'e}}, P.,
  {Storm}, J., \& {Bono}, G. 2004, \aj, 128, 1167

\bibitem[{{Grieve} {et~al.}(1985){Grieve}, {Madore}, \& {Welch}}]{Grieve85}
{Grieve}, G.~R., {Madore}, B.~F., \& {Welch}, D.~L. 1985, \apj, 294, 513

\bibitem[{{Hilditch} {et~al.}(2005){Hilditch}, {Howarth}, \&
  {Harries}}]{Hilditch05}
{Hilditch}, R.~W., {Howarth}, I.~D., \& {Harries}, T.~J. 2005, \mnras, 357, 304

\bibitem[{{Jha} {et~al.}(2007){Jha}, {Riess}, \& {Kirshner}}]{Jha07}
{Jha}, S., {Riess}, A.~G., \& {Kirshner}, R.~P. 2007, \apj, 659, 122

\bibitem[{{Kanbur} {et~al.}(2007){Kanbur}, {Ngeow}, {Nanthakumar}, \&
  {Stevens}}]{Kanbur07}
{Kanbur}, S.~M., {Ngeow}, C., {Nanthakumar}, A., \& {Stevens}, R. 2007, \pasp,
  119, 512

\bibitem[{{Keller} \& {Wood}(2006)}]{Keller06}
{Keller}, S.~C. \& {Wood}, P.~R. 2006, \apj, 642, 834

\bibitem[{{Kennicutt} {et~al.}(1998){Kennicutt}, {Stetson}, {Saha}, {Kelson},
  {Rawson}, {Sakai}, {Madore}, {Mould}, {Freedman}, {Bresolin}, {Ferrarese},
  {Ford}, {Gibson}, {Graham}, {Han}, {Harding}, {Hoessel}, {Huchra}, {Hughes},
  {Illingworth}, {Macri}, {Phelps}, {Silbermann}, {Turner}, \&
  {Wood}}]{Kennicutt98}
{Kennicutt}, Jr., R.~C., {Stetson}, P.~B., {Saha}, A., {Kelson}, D., {Rawson},
  D.~M., {Sakai}, S., {Madore}, B.~F., {Mould}, J.~R., {Freedman}, W.~L.,
  {Bresolin}, F., {Ferrarese}, L., {Ford}, H., {Gibson}, B.~K., {Graham},
  J.~A., {Han}, M., {Harding}, P., {Hoessel}, J.~G., {Huchra}, J.~P., {Hughes},
  S.~M.~G., {Illingworth}, G.~D., {Macri}, L.~M., {Phelps}, R.~L.,
  {Silbermann}, N.~A., {Turner}, A.~M., \& {Wood}, P.~R. 1998, \apj, 498, 181

\bibitem[{{Kochanek}(1997)}]{Kochanek97}
{Kochanek}, C.~S. 1997, \apj, 491, 13

\bibitem[{{Leavitt}(1908)}]{Leavitt1908}
{Leavitt}, H.~S. 1908, Annals of Harvard College Observatory, 60, 87

\bibitem[{{Lejeune} \& {Schaerer}(2001)}]{Lejeune01}
{Lejeune}, T. \& {Schaerer}, D. 2001, \aap, 366, 538

\bibitem[{{Macri}(2005)}]{Macri05}
{Macri}, L.~M. 2005, ArXiv Astrophysics e-prints, astro-ph/0507648

\bibitem[{{Macri} {et~al.}(2006){Macri}, {Stanek}, {Bersier}, {Greenhill}, \&
  {Reid}}]{Macri06}
{Macri}, L.~M., {Stanek}, K.~Z., {Bersier}, D., {Greenhill}, L.~J., \& {Reid},
  M.~J. 2006, \apj, 652, 1133

\bibitem[{{Madore}(1975)}]{Madore75}
{Madore}, B.~F. 1975, \apjs, 29, 219

\bibitem[{{Madore} \& {Freedman}(1991)}]{Madore91}
{Madore}, B.~F. \& {Freedman}, W.~L. 1991, \pasp, 103, 933

\bibitem[{{Mochejska} {et~al.}(2000){Mochejska}, {Macri}, {Sasselov}, \&
  {Stanek}}]{Mochejska00}
{Mochejska}, B.~J., {Macri}, L.~M., {Sasselov}, D.~D., \& {Stanek}, K.~Z. 2000,
  \aj, 120, 810

\bibitem[{{Moffett} {et~al.}(1998){Moffett}, {Gieren}, {Barnes}, \&
  {Gomez}}]{Moffett98}
{Moffett}, T.~J., {Gieren}, W.~P., {Barnes}, III, T.~G., \& {Gomez}, M. 1998,
  \apjs, 117, 135

\bibitem[{{Pagel} {et~al.}(1978){Pagel}, {Edmunds}, {Fosbury}, \&
  {Webster}}]{Pagel78}
{Pagel}, B.~E.~J., {Edmunds}, M.~G., {Fosbury}, R.~A.~E., \& {Webster}, B.~L.
  1978, \mnras, 184, 569

\bibitem[{{Payne-Gaposchkin} \& {Gaposchkin}(1966)}]{Payne66}
{Payne-Gaposchkin}, C. \& {Gaposchkin}, S. 1966, Smithsonian Contributions to
  Astrophysics, 9, 1

\bibitem[{{Peimbert} {et~al.}(2005){Peimbert}, {Peimbert}, \&
  {Ruiz}}]{Peimbert05}
{Peimbert}, A., {Peimbert}, M., \& {Ruiz}, M.~T. 2005, \apj, 634, 1056

\bibitem[{{Peimbert} \& {Torres-Peimbert}(1976)}]{Peimbert76}
{Peimbert}, M. \& {Torres-Peimbert}, S. 1976, \apj, 203, 581

\bibitem[{{Phillips}(1993)}]{Phillips93}
{Phillips}, M.~M. 1993, \apjl, 413, L105

\bibitem[{{Pietrukowicz}(2001)}]{Pietrukowicz01}
{Pietrukowicz}, P. 2001, Acta Astronomica, 51, 247

\bibitem[{{Pietrzy{\'n}ski} {et~al.}(2002){Pietrzy{\'n}ski}, {Gieren},
  {Fouqu{\'e}}, \& {Pont}}]{Pietrzynski02}
{Pietrzy{\'n}ski}, G., {Gieren}, W., {Fouqu{\'e}}, P., \& {Pont}, F. 2002, \aj,
  123, 789

\bibitem[{{Pietrzy{\'n}ski} {et~al.}(2006){Pietrzy{\'n}ski}, {Gieren},
  {Soszy{\'n}ski}, {Udalski}, {Bresolin}, {Kudritzki}, {Mennickent}, {Walker},
  {Garcia}, {Szewczyk}, {Szyma{\'n}ski}, {Kubiak}, \&
  {Wyrzykowski}}]{Pietrzynski06}
{Pietrzy{\'n}ski}, G., {Gieren}, W., {Soszy{\'n}ski}, I., {Udalski}, A.,
  {Bresolin}, F., {Kudritzki}, R.-P., {Mennickent}, R., {Walker}, A., {Garcia},
  A., {Szewczyk}, O., {Szyma{\'n}ski}, M., {Kubiak}, M., \& {Wyrzykowski},
  {\L}. 2006, \aj, 132, 2556

\bibitem[{{Pietrzy{\'n}ski} {et~al.}(2004){Pietrzy{\'n}ski}, {Gieren},
  {Udalski}, {Bresolin}, {Kudritzki}, {Soszy{\'n}ski}, {Szyma{\'n}ski}, \&
  {Kubiak}}]{Pietrzynski04}
{Pietrzy{\'n}ski}, G., {Gieren}, W., {Udalski}, A., {Bresolin}, F.,
  {Kudritzki}, R.-P., {Soszy{\'n}ski}, I., {Szyma{\'n}ski}, M., \& {Kubiak}, M.
  2004, \aj, 128, 2815

\bibitem[{{Pojmanski}(1997)}]{Pojmanski97}
{Pojmanski}, G. 1997, Acta Astronomica, 47, 467

\bibitem[{{Prieto} {et~al.}(2006){Prieto}, {Rest}, \& {Suntzeff}}]{Prieto06}
{Prieto}, J.~L., {Rest}, A., \& {Suntzeff}, N.~B. 2006, \apj, 647, 501

\bibitem[{{Riess} {et~al.}(2004){Riess}, {Strolger}, {Tonry}, {Casertano},
  {Ferguson}, {Mobasher}, {Challis}, {Filippenko}, {Jha}, {Li}, {Chornock},
  {Kirshner}, {Leibundgut}, {Dickinson}, {Livio}, {Giavalisco}, {Steidel},
  {Ben{\'{\i}}tez}, \& {Tsvetanov}}]{Riess04}
{Riess}, A.~G., {Strolger}, L.-G., {Tonry}, J., {Casertano}, S., {Ferguson},
  H.~C., {Mobasher}, B., {Challis}, P., {Filippenko}, A.~V., {Jha}, S., {Li},
  W., {Chornock}, R., {Kirshner}, R.~P., {Leibundgut}, B., {Dickinson}, M.,
  {Livio}, M., {Giavalisco}, M., {Steidel}, C.~C., {Ben{\'{\i}}tez}, T., \&
  {Tsvetanov}, Z. 2004, \apj, 607, 665

\bibitem[{{Romaniello} {et~al.}(2008){Romaniello}, {Primas}, {Mottini},
  {Pedicelli}, {Lemasle}, {Bono}, {Fran{\c c}ois}, {Groenewegen}, \&
  {Laney}}]{Romaniello08}
{Romaniello}, M., {Primas}, F., {Mottini}, M., {Pedicelli}, S., {Lemasle}, B.,
  {Bono}, G., {Fran{\c c}ois}, P., {Groenewegen}, M.~A.~T., \& {Laney}, C.~D.
  2008, \aap, 488, 731

\bibitem[{{Sandage} {et~al.}(2008){Sandage}, {Tammann}, \&
  {Reindl}}]{Sandage08}
{Sandage}, A., {Tammann}, G.~A., \& {Reindl}, B. 2008, ArXiv e-prints

\bibitem[{{Schlegel} {et~al.}(1998){Schlegel}, {Finkbeiner}, \&
  {Davis}}]{Schlegel98}
{Schlegel}, D.~J., {Finkbeiner}, D.~P., \& {Davis}, M. 1998, \apj, 500, 525

\bibitem[{{Schwarzenberg-Czerny}(1989)}]{Schwarzenberg89}
{Schwarzenberg-Czerny}, A. 1989, \mnras, 241, 153

\bibitem[{{Skillman} \& {Kennicutt}(1993)}]{Skillman93}
{Skillman}, E.~D. \& {Kennicutt}, Jr., R.~C. 1993, \apj, 411, 655

\bibitem[{{Spergel} {et~al.}(2003){Spergel}, {Verde}, {Peiris}, {Komatsu},
  {Nolta}, {Bennett}, {Halpern}, {Hinshaw}, {Jarosik}, {Kogut}, {Limon},
  {Meyer}, {Page}, {Tucker}, {Weiland}, {Wollack}, \& {Wright}}]{Spergel03}
{Spergel}, D.~N., {Verde}, L., {Peiris}, H.~V., {Komatsu}, E., {Nolta}, M.~R.,
  {Bennett}, C.~L., {Halpern}, M., {Hinshaw}, G., {Jarosik}, N., {Kogut}, A.,
  {Limon}, M., {Meyer}, S.~S., {Page}, L., {Tucker}, G.~S., {Weiland}, J.~L.,
  {Wollack}, E., \& {Wright}, E.~L. 2003, \apjs, 148, 175

\bibitem[{{Stanek} \& {Udalski}(1999)}]{Stanek99}
{Stanek}, K.~Z. \& {Udalski}, A. 1999, ArXiv Astrophysics e-prints,
  astro-ph/9909346

\bibitem[{{Tegmark} {et~al.}(2004){Tegmark}, {Strauss}, {Blanton}, {Abazajian},
  {Dodelson}, {Sandvik}, {Wang}, {Weinberg}, {Zehavi}, {Bahcall}, {Hoyle},
  {Schlegel}, {Scoccimarro}, {Vogeley}, {Berlind}, {Budavari}, {Connolly},
  {Eisenstein}, {Finkbeiner}, {Frieman}, {Gunn}, {Hui}, {Jain}, {Johnston},
  {Kent}, {Lin}, {Nakajima}, {Nichol}, {Ostriker}, {Pope}, {Scranton},
  {Seljak}, {Sheth}, {Stebbins}, {Szalay}, {Szapudi}, {Xu}, {Annis},
  {Brinkmann}, {Burles}, {Castander}, {Csabai}, {Loveday}, {Doi}, {Fukugita},
  {Gillespie}, {Hennessy}, {Hogg}, {Ivezi{\'c}}, {Knapp}, {Lamb}, {Lee},
  {Lupton}, {McKay}, {Kunszt}, {Munn}, {O'Connell}, {Peoples}, {Pier},
  {Richmond}, {Rockosi}, {Schneider}, {Stoughton}, {Tucker}, {vanden Berk},
  {Yanny}, \& {York}}]{Tegmark04}
{Tegmark}, M., {Strauss}, M.~A., {Blanton}, M.~R., {Abazajian}, K., {Dodelson},
  S., {Sandvik}, H., {Wang}, X., {Weinberg}, D.~H., {Zehavi}, I., {Bahcall},
  N.~A., {Hoyle}, F., {Schlegel}, D., {Scoccimarro}, R., {Vogeley}, M.~S.,
  {Berlind}, A., {Budavari}, T., {Connolly}, A., {Eisenstein}, D.~J.,
  {Finkbeiner}, D., {Frieman}, J.~A., {Gunn}, J.~E., {Hui}, L., {Jain}, B.,
  {Johnston}, D., {Kent}, S., {Lin}, H., {Nakajima}, R., {Nichol}, R.~C.,
  {Ostriker}, J.~P., {Pope}, A., {Scranton}, R., {Seljak}, U., {Sheth}, R.~K.,
  {Stebbins}, A., {Szalay}, A.~S., {Szapudi}, I., {Xu}, Y., {Annis}, J.,
  {Brinkmann}, J., {Burles}, S., {Castander}, F.~J., {Csabai}, I., {Loveday},
  J., {Doi}, M., {Fukugita}, M., {Gillespie}, B., {Hennessy}, G., {Hogg},
  D.~W., {Ivezi{\'c}}, {\v Z}., {Knapp}, G.~R., {Lamb}, D.~Q., {Lee}, B.~C.,
  {Lupton}, R.~H., {McKay}, T.~A., {Kunszt}, P., {Munn}, J.~A., {O'Connell},
  L., {Peoples}, J., {Pier}, J.~R., {Richmond}, M., {Rockosi}, C., {Schneider},
  D.~P., {Stoughton}, C., {Tucker}, D.~L., {vanden Berk}, D.~E., {Yanny}, B.,
  \& {York}, D.~G. 2004, \prd, 69, 103501

\bibitem[{{T{\"u}llmann} {et~al.}(2003){T{\"u}llmann}, {Rosa}, {Elwert},
  {Bomans}, {Ferguson}, \& {Dettmar}}]{Tullmann03}
{T{\"u}llmann}, R., {Rosa}, M.~R., {Elwert}, T., {Bomans}, D.~J., {Ferguson},
  A.~M.~N., \& {Dettmar}, R.-J. 2003, \aap, 412, 69

\bibitem[{{Udalski} {et~al.}(1999){Udalski}, {Szymanski}, {Kubiak},
  {Pietrzynski}, {Soszynski}, {Wozniak}, \& {Zebrun}}]{Udalski99}
{Udalski}, A., {Szymanski}, M., {Kubiak}, M., {Pietrzynski}, G., {Soszynski},
  I., {Wozniak}, P., \& {Zebrun}, K. 1999, Acta Astronomica, 49, 201

\bibitem[{{Urbaneja} {et~al.}(2005){Urbaneja}, {Herrero}, {Bresolin},
  {Kudritzki}, {Gieren}, {Puls}, {Przybilla}, {Najarro}, \&
  {Pietrzy{\'n}ski}}]{Urbaneja05}
{Urbaneja}, M.~A., {Herrero}, A., {Bresolin}, F., {Kudritzki}, R.-P., {Gieren},
  W., {Puls}, J., {Przybilla}, N., {Najarro}, F., \& {Pietrzy{\'n}ski}, G.
  2005, \apj, 622, 862

\bibitem[{{van Genderen}(1983)}]{vanGenderen83}
{van Genderen}, A.~M. 1983, \aaps, 52, 423

\end{thebibliography}

\end{document}